\documentclass[twocolumn]{autart}    

\usepackage {amsmath}
\usepackage {units} 
\usepackage {amssymb}
\usepackage{graphicx}          
                               
\def \R {\mathbb R}     

\begin{document}

\begin{frontmatter}
\runtitle{Structure of SPS confidence sets}  

\title{Perturbed Datasets Methods for Hypothesis Testing and Structure of Corresponding Confidence Sets \thanksref{footnoteinfo}} 

\thanks[footnoteinfo]{This paper was not presented at any IFAC 
meeting. Corresponding author S.~Kolumb\'{a}n. Tel. +36-1-463-2870. 
Fax +36-1-463-2871.}

\author[Aut]{S\'{a}ndor Kolumb\'{a}n}\ead{Sandor.Kolumban@aut.bme.hu},    
\author[Aut]{Istv\'{a}n Vajk}\ead{Istvan.Vajk@aut.bme.hu},
\author[Elec]{Johan Schoukens}\ead{Johan.Schoukens@vub.ac.be}               

\address[Aut]{Budapest University of Technology and Economics, Department of Automation and Applied Informatics, Magyar tud\'{o}sok krt. 2 QB207, 1117 Budapest, Hungary \\ MTA-BME Control Engineering Research Group}  
\address[Elec]{Vrije Universiteit Brussel, Department ELEC, Pleinlaan 2, B1050 Brussels, Belgium}             


\begin{abstract}                          
Hypothesis testing methods that do not rely on exact distribution assumptions have been emerging lately. The method of sign-perturbed sums (SPS) is capable of characterizing confidence regions with exact confidence levels for linear regression and linear dynamical systems parameter estimation problems if the noise distribution is symmetric. This paper describes a general family of hypothesis testing methods that have an exact user chosen confidence level based on finite sample count and without relying on an assumed noise distribution. It is shown that the SPS method belongs to this family and we provide another hypothesis test for the case where the symmetry assumption is replaced with exchangeability. In the case of linear regression problems it is shown that the confidence regions are connected, bounded and possibly non-convex sets in both cases. To highlight the importance of understanding the structure of confidence regions corresponding to such hypothesis tests it is shown that confidence sets for linear dynamical systems parameter estimates generated using the SPS method can have non-connected parts, which have far reaching consequences.
\end{abstract}

\end{frontmatter}

\section{Introduction}

We use models to describe a wide range of systems and phenomena. Such models can be derived from prior knowledge or inferred from measurement data. Parameter estimation aims at extracting models from noisy measurement data under the assumption that the data was generated by a process from a considered model class and measurements are contaminated with noise. Because of the noise, the extracted model will not match the nominal one that generated the data, but it should be close to it in some sense. When estimation is carried out, quantitative information should also be delivered along with the extracted model, describing the reliability of the model.

In the case of linear systems parameter estimation we know that the distribution of parameters, estimated in the least squares sense, converges to a Gaussian distribution under some mild moment conditions on the noise and an approximate confidence region can be constructed based on the limiting distribution. Of course, the central limit theorem is in the background of this result. When the amount of data is relatively large considering the noise levels and the number of estimated parameters, then the delivered point estimates are rather precise, the confidence regions are small and they have an approximately correct confidence level. However, for small datasets, all of this breaks down; the sample count is not large enough for the central limit theorem to take effect. The Gaussian distribution poorly approximates the distribution of the estimated parameters and the generated confidence ellipsoids can have a real confidence level arbitrarily far from the required one (\cite{Garatti2004}, \cite{ljung1999SysId}).

There is a growing interest for developing methods that do not rely on the central limit theorem or on Gaussian assumptions about the noise (\cite{Campi2005NonAssID}, \cite{Campi2010NonAssLinearTF}, \cite{Csaji2012Sysid}, \cite{Csaji2012CDC}, \cite{Dalai2007LSCRNonLin}). The $16^{\textrm{th}}$ IFAC Symposium on System Identification had a plenary session dedicated to this topic \cite{Campi2012Plenary}.  These procedures provide means to test a given model, whether it is the nominal one or not, accepting the nominal model with user given exact probability. Seldom do these provide a description of the whole generated confidence set. This is a problem when the whole confidence region needs to be visualized or evaluated.
`
The current paper describes a wide family of hypothesis testing algorithms (methods of perturbed datasets) that are not relying on conservative distribution assumptions. We show that the method of sign-perturbed sums (SPS), introduced in \cite{Csaji2012Sysid} and \cite{Csaji2012CDC}, belongs to this family. These methods provide a hypothesis test for parameters of linear regression problems, linear dynamical systems and other non-linear models. The exact confidence level of the SPS hypothesis test was proven in the previously cited papers. Our interest is in generalizing this algorithm and characterizing the structure of confidence sets whose characteristic function is the hypothesis test. In order to do this, we formulate the steps of the SPS method in the general framework. This formulation will provide helpful insight into the nature of perturbed datasets methods.

Perturbed dataset methods generate confidence sets for arbitrary confidence level based on confidence sets corresponding to confidence level $\nicefrac{1}{2}$. Properties of these confidence sets are inherited by the derived confidence sets with other confidence levels. For this reason, we examine the properties of the $\nicefrac{1}{2}$ confidence sets for both linear regression problems and linear dynamical systems. We present the conditions required in order to have bounded confidence sets in the linear regression case. In the dynamical systems case we show that the confidence regions can be disconnected, pointing out the fact that careful analysis of the structure of confidence sets is needed.

Our contribution is twofold. The first major contribution of the paper is that, in the footsteps of the SPS method, we describe the family of perturbed datasets hypothesis testing methods, which are distribution-free  having exact user prescribed confidence level. We describe a perturbed dataset hypothesis test that replaces the symmetry assumption of the SPS method with exchangeability \cite{Aldous1985}.  Our second contribution is the precise structural analysis of the confidence regions corresponding to both this algorithm and the SPS hypothesis test. While for linear regression problems these confidence sets are connected and bounded, for dynamical systems parameter estimation problems they can become non-connected.

The structure of the paper is as follows. We describe the framework of perturbed datasets methods in Section~\ref{sec:GenericView}, where we also show how the SPS method fits into this framework and what its building blocks are. We highlight possible modifications in order to create other hypothesis testing algorithms. In Section~\ref{sec:BENonSym} we illustrate the potential of the framework by constructing exact connected confidence sets for linear regression problems without the symmetry assumption. We illustrate in Section~\ref{sec:DynSys} that this question is more complex for parameter estimates of linear dynamical systems by showing that the SPS method might result in non-connected confidence sets. Concluding remarks are given in Section~\ref{sec:Conc}.

\section{Perturbed Datasets}
\label{sec:GenericView}

The goal of this section is to present a new general framework for hypothesis testing methods similar to the SPS method. We suppose that the measurements come from a model
\begin{equation}
Y = f(\theta^*, X, N)
\end{equation}
where $f$ is a known mapping, $\theta^* \in \R^{n_\theta}$ contains the unknown parameters of the model, $X$ contains the measured input, $N$ contains the not measured contaminating noise and $Y$ contains the measured output values. Please note, that the dimensions of $X$, $Y$ and $N$ depend on the estimation problem at hand.

\begin{assum}[Invertibility with respect to noise] Based on the selected model parameters $\theta$ and the measurement values $X$ and $Y$, a corresponding noise realization is always uniquely determined: $\exists \ f^*: \Theta \times  \mathcal{X} \times \mathcal{Y} \rightarrow \mathcal{N}$
such that 
\[
Y = f(\theta, X, N) \Rightarrow N = f^*(\theta, X, Y)
\]
\end{assum}
For ease of notation we introduce the short hand notation $N(\theta) = f^*(\theta, X, Y)$.

We want to construct a hypothesis test for a parameter vector $\theta$ based on values of $X$ and $Y$.

Let us introduce the common notation $D$ to denote the input and output dataset. The task is to generate a hypothesis test for a parameter vector $\theta$ without exact knowledge about the distribution of the disturbances $N$. Exact knowledge about the distribution of a random variable is needed in order to generate confidence regions. This random variable will be the ordering of $m$ independent and identically distributed random variables $Z_i$, $i=1,\ldots, m$ defined on an appropriately chosen probability space.

The extra randomness needed to create the variables $Z_i$ is given in a data perturbation setup $\Gamma$.

The abstract steps of the method family are as follows.
\begin{enumerate}
\item[i.] \label{en:BEi} Generate $m$ different datasets $D^{(i)}(D,\theta)$ based on a random setup $\Gamma$.
\item[ii.] \label{en:BEii} Define a performance measure $Z$ and define $Z_i$ as the performance of model $\theta$ on the dataset $D^{(i)}(D,\theta)$.
\item[iii.] \label{en:BEiii} Create the random variable $O$ which is the (well defined) ordering of the values $Z_i$.
\item[iv.]  \label{en:BEiv} Define the subset of the $m!$ possible orderings in which the model $\theta$ is accepted.
\end{enumerate}

If the procedure used to create the datasets $D^{(i)}(D,\theta)$ is such that there exists a sigma-algebra $\sigma$ for which these datasets are conditionally independent and identically distributed conditioned on $\theta = \theta^*$ and $\sigma$, then the values $Z_i$ will also be conditionally independent and identically distributed. If equality between  $Z_i$ values occurs with zero probability then every ordering will be equally probable with probability $\nicefrac{1}{m!}$. This allows setting the confidence level of the confidence set by selecting the appropriate number of orderings.

The rest of the section consists of two parts. First, we provide the building blocks of the SPS method in the perturbed datasets framework (Section~\ref{sec:BESPS}). This helps to understand the role of these building blocks in the framework. We continue in Section~\ref{sec:BEGeneralConsideration} with some general notes to offer further insight into the capabilities of the framework.

\subsection{Building blocks of the SPS method}
\label{sec:BESPS}

In this section we go through the building blocks of the presented general framework using the SPS method. The first of these is the procedure used to generate different datasets that are conditionally independent and identically distributed. The second building block is the performance measure that is used to evaluate the given model on these generated datasets. The third one is the definition of the ordering and the last building block is the set of accepted orderings.

In case of the SPS method, the data perturbation setup $\Gamma$ consists of
\begin{enumerate}
\item[a)] $m$ sign sequences $\alpha_\Gamma(i,k)$, $i=1,\ldots, m$, $k=1,\ldots,n$ where $\alpha_\Gamma(1,\cdot) = 1$ and $\alpha_\Gamma(i,\cdot)=\pm1$ with equal probability, independently of everything else, for $i=2,\ldots,m$. 
\item[b)] A random permutation $\pi$ of the numbers $1,\ldots, m$. Each permutation is selected with probability $1/m!$ and independently of everything else.
\end{enumerate}
 
\textit{i. Generating the perturbed datasets}: If $\theta=\theta^*$ then the noise realization $N(\theta) = N(\theta^*)$ is the actual random noise that contaminated the measurements. The SPS method assumes that the noise has symmetric distribution around zero. If the sign of these noise values is changed using a random sign sequence 
$\alpha_\Gamma(i,k)$ then equally probable noise sequences are generated.

Let $W_i\in \R^{n \times n}$ be defined as the diagonal matrix containing the signs $\alpha_\Gamma(i,k)$ and let the perturbed noise sequences be defined as
\begin{equation}
N^{(i)}(\theta, \Gamma) = W_i N(\theta)
\end{equation}
Note that $N^{(1)}(\theta, \Gamma) = N(\theta)$ and $N(\theta^*) = N$.

Perturbed datasets can be created using the perturbed noise realizations as
\begin{equation}
Y^{(i)} = f(\theta, X, N^{(i)}) \quad i = 1,\ldots , m 
\end{equation}
 
This means that $m$ independent sign sequences define $m$ conditionally independent noise sequences. The sigma-algebra $\sigma$ is generated by the input values $X$ and the absolute values of the real noise $|N_k|$. It is a technical but important detail that if one of the sign sequences is chosen to be all-one then the conditional independence conditioned on $\sigma$ still holds. So the $m$ different datasets in the case of the SPS method are generated using random signs. $D^{(i)}(D,\theta)$ contains the input $X$ and the perturbed outputs $Y^{(i)}$.

\textit{ii. The performance measure} that is used to evaluate the model $\theta$ on the generated datasets: In the original publications \cite{Csaji2012Sysid} \cite{Csaji2012CDC} this is chosen to be some weighted norm of the gradient of the quadratic cost function at model $\tilde{\theta} =  \theta$. The cost function is
\begin{equation}
J^{(i)}_\theta(\tilde{\theta}) = \frac{1}{N} \left[f^*\left(\tilde{\theta}, X, Y^{(i)}\right)\right]^T f^*\left(\tilde{\theta}, X, Y^{(i)}\right)
\end{equation}
Either the norm of the gradient is taken as it is, or the estimated covariance matrix is used as the weighting matrix \cite{Csaji2012Sysid}
\begin{equation}
Z_i(\theta, \Gamma) = \left\|\frac{\partial J_\theta^{(i)}(\theta)}{\partial \theta}\right\|^2_S
\end{equation}
where $S$ is either the identity matrix or the estimate of the covariance matrix belonging to the original problem.

\textit{iii. Creation of the random ordering $O$:} As the random sign sequences are discrete random variables, there can be equal values $Z_i$ with non zero probability. In case of the SPS method the ordering is defined as the order of the indices corresponding to the decreasing ordering of the values $Z_i$. If $Z_i =Z_j$ for some $i \neq j$, so the ordering is not uniquely defined, then their relation is defined by the position of $i$ and $j$ in the random permutation $\pi$ given in the setup $\Gamma$.

We note that the original papers (\cite{Csaji2012Sysid} \cite{Csaji2012CDC}) describe a different procedure to resolve ties but it is not difficult to show that the two tie resolution schemes are equivalent.

\textit{iv. Accepted orderings}: Every possible outcome of the random variable $O$ has probability $1/m!$. For a hypothesis test with confidence level $q = k/m!$ we have to select $k$ different permutations as accepted permutations. If $O$ turns out to be one of these accepted permutations than the test accepts $\theta$.

The set of accepted orderings by the SPS method is determined by the rank of $Z_1$, that is the position of $1$ in the ordering $O$. Since only the position of $Z_1$ is used, the resolution of the confidence levels is only $1/m$. This means that confidence regions with confidence level $1-1/m$, $1-2/m$, $\ldots$, $\ldots$, $1-(m-1)/m$ are generated. We note that the resolution can be increased from $1/m$ to $1/m!$ by selecting the accepted orderings individually not just based on the rank of $1$.

\subsection{General considerations for the building blocks}
\label{sec:BEGeneralConsideration}

This section tries to offer some insight into the possibilities and restrictions offered by hypothesis testing methods belonging to the perturbed datasets class. We go through the four building blocks one-by-one commenting on them.

\textit{i. Generating the perturbed datasets:} Randomness in the data comes from the noise realization that contaminates the measurements. In order to generate identically distributed datasets, a perturbation procedure is needed that leaves the joint distribution of the noise sequence invariant. The more assumptions we have about the noise, the more possibilities are available for these transforms. In the case of the SPS method symmetry is assumed, thus any sign-perturbation is a suitable transformation. If it is assumed that the noise samples are exchangeable random variables \cite{Aldous1985} (i.i.d. for example) then permutations are suitable transformations. Based on this observation we present a method in Section~\ref{sec:BENonSym} which replaces the symmetry assumption of SPS with exchangeability.  If both symmetry and exchangeability are assumed then sign-perturbed permutations can be used. The class of perturbation methods to create identically distributed datasets grows with the amount of assumptions. 

\textit{ii. The performance measure} used to evaluate the given model $\theta$ on the generated perturbed datasets presents an inner controversy of all perturbed datasets methods. We illustrate this controversy on the SPS method, but it is easy to see that the same issue is present for every method belonging to this class.

\begin{rem} In order to produce meaningful random variables $Z_i$, the selected performance measure $Z$ should be such that it is \textbf{not} invariant under the perturbing transformation used to create the perturbed datasets.
\end{rem}

The SPS method assumes symmetry of the noise. When a point estimate is sought, in such a case, the sensible choices of cost functions to be minimized are symmetric. If we think of the least squares method or the prediction error method, these methods minimize a quadratic function of the errors. Although different cost functions $J^{(i)}_\theta (\cdot)$ correspond to each perturbed dataset, these cost functions have the same value at model $\theta$ that was used to generate the datasets.
\begin{equation}
J^{(i)}_\theta (\theta) = J^{(j)}_\theta (\theta) \qquad \forall i, j = 1,\ldots, m
\end{equation}
Of course, the identity $J^{(i)}_\theta (\theta') = J^{(j)}_\theta (\theta')$ will usually not hold if $\theta' \neq \theta$.

This controversy is inherently part of the perturbed datasets framework. 

\textit{iii. Creation of the random ordering $O$}: If the performance measure and the perturbed datasets guarantee that $\mathbb{P}\left(Z_i=Z_j\right) = 0$ for $i \neq j$ then simple ordering of the $Z_i$ values is sufficient. Otherwise a tie resolution is needed similar to what is shown earlier in the case of the SPS method.

\textit{iv. Accepted orderings}: Having discussed the other building blocks, we turn our attention to the selection of appropriate orderings of the $Z_i$ variables that we accept. If the order of the $Z_i(D, \theta)$ variables turns out to be from the selected ones, we accept the model $\theta$ to be in the confidence set. Finding a distribution invariant transformation and a performance measure that is not invariant under this transform is usually easy. If we only aim for the confidence level of the created hypothesis test then we can select the acceptable orderings any way we want. If any $r=m!-q(m-1)!$ orderings are selected, then the confidence level of the corresponding hypothesis test will be $r/m! = 1-q/m$.

If we think about confidence regions for the expected value of Gaussian random variables with known variance but unknown mean value, an infinite choice of confidence regions can be constructed. However, we prefer symmetric confidence regions centred around the average instead of confidence regions of the form $(-\infty, a] \cup [b, \infty)$. The difficult part of all perturbed dataset methods is the choice of the accepted permutations in a way that the corresponding confidence regions will be useful.

Among others, hypothesis tests are used in two fundamental ways. The first one is when a particular hypothesis is tested and a yes or no answer is expected about the acceptance. The second one is when the entire confidence set needs to be evaluated somehow. It is important that when we judge the level of uncertainty based on a given set then we should have guarantees that no point outside that set will pass the hypothesis test. Interval analytic methods can be used to find arbitrary fine approximations of the confidence set inside a given initial box (as given in \cite{Kieffer2014507} for the linear regression SPS method). If this initial box does not contain all components of the confidence region then the resulting approximation will contain no information about this fact, resulting in a bad approximation of the confidence set. This is why rigorous analysis of the structure of confidence sets belonging to a given hypothesis test is so important.

\section{Dropping the symmetry condition}
\label{sec:BENonSym}

The SPS method constructs confidence regions for models assuming that the disturbing noise samples have symmetric distribution, not necessary identical. We present a different method that belongs to the class of perturbed datasets methods that handles the case where the disturbing noise is an exchangeable sequence of random variables. Independent and identically distributed variables belong to this class. Note that the symmetry condition is not needed; it is replaced with exchangeability. We only present the method for the linear regression problem, but it generalizes in a straightforward way to dynamical systems as well.

The linear regression problem in this situation can be formalized as 
\begin{equation}
\label{eq:LinRegProblemDef2}
Y = X^T \theta^*  + N
\end{equation}
where $X\in \R^{n_\theta \times n}$ is the matrix of regressors, $Y\in \R^n$ is the vector of observations and $N \in \R^n$ is a vector of independent and identically distributed random variables. It should be emphasized that no moment, symmetry or centrality conditions are imposed on this distribution.

We consider the linear regression problem (\ref{eq:LinRegProblemDef2}) with the assumption that the noise sequence $N$ is a sequence of exchangeable random variables. Based on this assumption we construct bounded connected confidence regions for parameter $\theta$. Note, that no moment or symmetry conditions are imposed on the noise sequence.

\subsection{Perturbed dataset building blocks}
\label{sec:PertDSBB}

The SPS method used $m$ random sign sequences to create the perturbed noise realizations. In case of exchangeable noise sequences we use $m$ random permutations of the noise samples to perturb the data. These $m$ permutations for generating the noise realizations are denoted by $\pi_i$, $i=1, \ldots, m$. $\pi_1$ is chosen to be the identity permutation, the others are uniformly selected, independently of everything else. Let $P_i$ denote the permutation matrix corresponding to the permutation $\pi_i$, so $P_1$ is the $n$ dimensional identity matrix.

The performance measure of model $\theta$ on the perturbed dataset $D^{(i)}(D,\theta)$ will be a weighted distance between $\theta$ and the least squares estimate corresponding to $D^{(i)}(D,\theta)$.

If $\theta^{(i)}$ denotes the least squares estimate corresponding to $D^{(i)}(D,\theta)$, then
\begin{eqnarray}
\theta^{(i)} &=& \left[X X^T \right]^{-1} X \hat{Y}^{(i)} = \\
 &=& \left[X X^T \right]^{-1} X (X^T \theta + P_i(Y - X^T \theta )) = \\
 &=& \theta + \left[X X^T \right]^{-1} X  P_i(Y - X^T \theta )
\end{eqnarray}
If the performance measure is defined as the weighted distance between $\theta$ and $\theta^{(i)}$ with weighting matrix $X X^T$ then it can be written as
\[
Z_i(\theta, \Gamma) = (Y - X^T \theta )^T P_i^T X^T \left[X X^T \right]^{-1} X P_i (Y - X^T \theta )
\]

Note that this weighting is a natural choice as  $X X^T$ is the inverse of the estimated covariance matrix corresponding to the estimate $\theta^{(i)}$.

The ordering $O$ and the accepted permutations for the model $\theta$ are determined the same way as it is done for the SPS method. For the prescribed confidence level $r/m!$, $r$ permutations are chosen such that the position of $1$ in them is as big as possible.

\subsection{Structure of confidence sets}

This section contains the structural analysis of the confidence regions corresponding to the hypothesis test defined in the previous section.

\begin{defn}[Sufficiently exciting input] We say that the problem input $X$ is sufficiently exciting with respect to a permutation matrix $P$ if $Q > 0$ holds, where
\[
Q = X X^T - X P^T X^T \left[X X^T \right]^{-1} X P X^T
\]
\end{defn}

For the input $X$ to be exciting enough is more restrictive than in usual linear regression problems. The constant input $X_k = 1$, $k=1,\ldots,n$ is not sufficiently exciting for any permutation matrix as $Q=0$ in every case. There are no permutations that can sufficiently mix this input matrix. From the perspective of the permutations there is no input that is sufficiently exciting with respect to the identity permutation. This condition can be interpreted as the input $X$ should be sufficiently exciting in the regular sense ($XX^T$ is invertible), but additionally it also required that mixing the regressors along the time axes should result in a significantly different excitation.

\begin{thm}
\label{thm:BoundedPCSLR}
Let the perturbed noise sequences $N^{(i)}(\theta)$ be generated as
\[
N^{(i)}(\theta, \Gamma) = P_i N (\theta) = P_i (Y-X^T \theta)
\]
and the performance measure be the weighted norm 
\[
Z_i(\theta, \Gamma) =  (Y - X^T \theta )^T P_i^T X^T \left[X X^T \right]^{-1} X P_i (Y - X^T \theta )
\]
Out of the $m!$ possible permutations let the $r$ acceptable permutations be chosen in decreasing order of the position of $1$ until $r/m! = \alpha$.

Under these conditions, the confidence regions characterized by the corresponding perturbed dataset method are connected, containing the least squares estimate. If the input $X$ is sufficiently exciting with respect to the $m$ permutations then the confidence regions are also bounded.
\end{thm}
Before we prove the theorem, we formalize the randomization property of permutations. The statement of Lemma~\ref{lem:RandomPerm} immediately follows from the definition of independence of random variables.
\begin{lem}
\label{lem:RandomPerm}
Let $\pi_1$ be a fixed permutation and $\pi_2$ be a uniformly chosen random permutation. Let $\pi_3 = \pi_1 \pi_2$ be the permutation obtained by first applying $\pi_1$ and then $\pi_2$. Under these assumptions $\pi_3 = \pi_1 \pi_2$ is also a uniform random permutation and it is independent from $\pi_1$.
\end{lem}

\begin{pf*}{Proof.}
The proof of the exact confidence level goes the same way as it is done for the SPS method in \cite{Csaji2012CDC}. The only difference is that in the case of SPS a key element of the proof is the randomization property of random signs. This is exchanged with the randomization property of random permutations given in Lemma~\ref{lem:RandomPerm}.

In order to show that the characterized confidence regions are bounded and connected we prove this for confidence regions with confidence level $\nicefrac{1}{2}$. Sets for general confidence level are created as a union of intersections of such sets, preserving this structure.
We need to prove that $Z_1$ will outgrow $Z_i$ as $\|\theta^*-\theta\| \rightarrow \infty$ and we do this by showing that the difference $Z_1-Z_i\rightarrow \infty$. The values $Z_i$ can be written as
\begin{eqnarray*}
Z_i&=& \left(Y -X^T \theta^* \right)^T P_i^T X^T \left[X X^T \right]^{-1} X P_i \left(Y - X^T \theta^* \right) +\\
&+& 2 \left(Y -X^T \theta^* \right)^T P_i^T X^T \left[X X^T \right]^{-1} X P_i  X^T (\theta^* - \theta) + \\
&+& (\theta^* - \theta)^T X P_i^T X^T \left[X X^T \right]^{-1} X P_i X^T (\theta^* - \theta)
\end{eqnarray*}

This means that the limit $Z_1-Z_i \rightarrow \infty$ as $\|\theta^*-\theta\| \rightarrow \infty$ holds if $Q > 0$, where $Q$ is defined as
\begin{equation}
\label{eq:QPerm}
Q = X X^T - X P_i^T X^T \left[X X^T \right]^{-1} X P_i X^T
\end{equation}
Using the fact that $P_iP_i^T = I$ for all permutation matrices
\begin{equation}
Q = X \left( I - P_i^T X^T \left[X P_iP_i^T X^T \right]^{-1} X P_i \right) X^T
\end{equation}

The term in the middle is the difference of the identity matrix and a projection matrix defined by $P_i^T X^T$ showing that the eigenvalues of the middle term are $0$ or $1$. This shows that $Q \geq 0$. If the input is sufficiently exciting with respect to $P_i$ then $Q>0$ also holds.

As the level $1/2$ confidence set is characterized by a lower level set of a convex quadratic function it is always connected and convex.

The least squares estimate is always contained in the confidence region as $Z_1(\theta^{LS}, \Gamma) = 0$, thus it is always smaller or equal than the other $Z_i$ values. \qed
\end{pf*}

We note that by appropriately defining sufficient excitation with respect to a random sign sequence a theorem similar to Theorem~\ref{thm:BoundedPCSLR} can be proven for the SPS method as well (both with and without using the weighting matrix in the performance measure).

\section{Dynamical Systems with SPS}
\label{sec:DynSys}

In the previous section we focused our attention on the structure of confidence sets generated for linear regression problems. We carry out similar analysis of the method for dynamical systems. A negative result is presented showing that in case of dynamical systems the SPS method characterizes non-connected confidence regions. 

Using the derivative of the quadratic cost function as performance measure proved to be useful in the linear regression case. That was mainly due to the convexity of the cost function. We illustrate what kind of problems can appear if the gradient of a more complex cost function is used as performance measure. Namely, the generated confidence regions can become disconnected.

Let us present a situation where the SPS method produces a non-connected confidence region. The considered problem is a two parameter output error problem \cite{ljung1999SysId}. The problem is defined with nominal system
\begin{equation}
G(z^{-1}, \theta) = \frac{\theta_1 z^{-1} }{1 + \theta_2 z^{-1}} \qquad \theta = \left[0.9 \ -0.1\right]^T
\end{equation}
unit step input starting at $t=0$ for $n=7$ samples and noise values
\begin{equation}
N = 10^{-2}\left[ \begin{array}{ccccccc}
-2.1 & -0.8 & -0.3 & -0.4 & 1.0 & 0.7 & 1.5
\end{array}  \right]^T
\end{equation}
The used SPS setup is a $\nicefrac{1}{2}$ confidence level one, with the second sign sequence
\begin{equation}
\alpha(2, \cdot) = \left[  \begin{array}{ccccccc}
1 & -1 & 1 & -1 & 1 & 1 & -1
\end{array}  \right]^T
\end{equation}
and tie order $\pi = [1 \ 2]$.

\begin{figure}[htb]
 \centerline{\includegraphics[width=0.9\columnwidth]{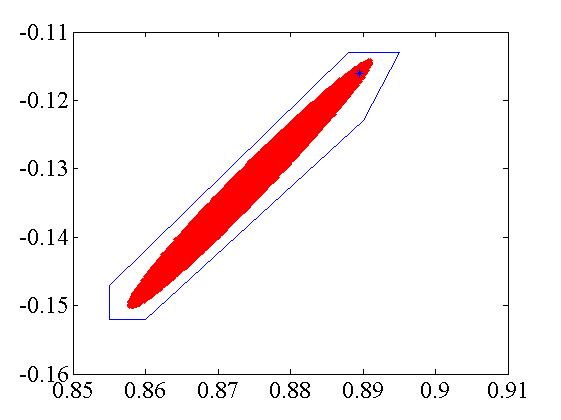}}
 \caption{Connected component of the confidence region around the pem estimate.}
 \label{fig:AroundPEM}
\end{figure}

Fig.~\ref{fig:AroundPEM} shows the connected component of the confidence region around the prediction error estimate. The interior of the marked polygon was tested for membership. It is really tempting to think that the located region is the entire confidence region. 

\begin{figure}[htb]
 \centerline{\includegraphics[width=0.9\columnwidth]{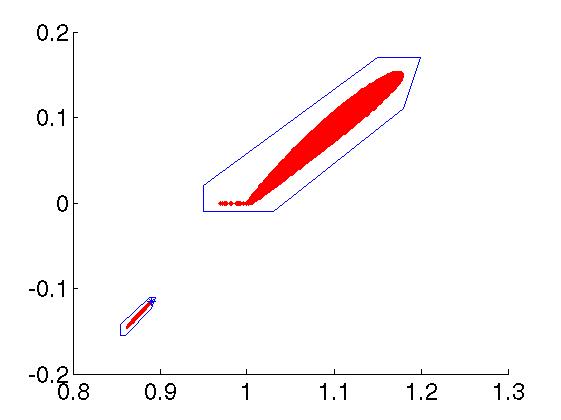}}
 \caption{Two connected components of the confidence set.}
 \label{fig:United}
\end{figure}

It is easy to show that the quadratic prediction error cost function has an inflection point on the line $\theta_2=0$. By finding this inflection point and checking its neighbourhood, another connected component of the confidence region can be discovered. The two found connected components of the confidence set are presented in Fig.~\ref{fig:United}.

If we know nothing about the distribution of the noise, this second connected component cannot be discarded. Also the  models from this second component has nothing to do with the nominal model. This is due to the fact that every zero of the gradient of the cost function will be included in the confidence set (local extrema and inflection points).

This example shows that confidence regions for dynamical systems generated using the SPS method can be disconnected, this is an observations that has far reaching consequences.

First and foremost, we have no guaranties that there are no other components left undiscovered. Questions about the volume of the confidence set cannot be answered without guaranties that the entire set is discovered.

A confidence region should be a concise description of the possible models so it can be used later on without the entire data record. Without the whole original data record we are not able to say that the big area around the inflection point is negligible. Exploring only the connected component around the prediction error estimate will result in a set with confidence level less than what was prescribed. Suppose that the noise realization is drawn from Gaussian distribution with variance $0.0004$. The total probability of the noise realizations corresponding the models in the connected region around the inflection point is negligible. This can only be seen if the original dataset is still available.

\section{Concluding Remarks}
\label{sec:Conc}

We have presented a general framework that can be used to generate hypothesis testing methods with exact probability. Depending on prior assumptions about the noise distribution, different methods can be defined. If this prior assumption is symmetry, then the SPS method fits into the presented frame. To illustrate how other assumptions can be used, we presented a method that relies on the exchangeability property of the noise distribution.

In general, it is not difficult to create hypotheses tests in the presented framework that have a user prescribed exact confidence level. The challenging part is to ensure that the created method will result in confidence regions that satisfy our needs. We illustrated that the magnitude of the gradient is not a good choice as a performance measure when the whole set needs to be discovered, as it generates disconnected confidence regions. If we want to create connected approximate confidence regions for dynamical systems with the SPS method, the connected component around the prediction error estimate might be a good choice. It will have lower confidence than the prescribed level, but usually the difference will be negligible. For short data records this difference might be significantly smaller than what would be caused by using an asymptotic confidence region.

As we used it in Section~\ref{sec:BENonSym}, the appropriately weighted norm of the difference between the model $\theta$ and the minimizer $\theta^{(i)}$ of the cost function $J^{(i)}_\theta (\cdot)$ corresponding to the perturbed dataset is also a possible choice to measure model performance. In the linear regression case this can always be transformed into a measure using the derivative of the cost function, but this is no longer true when the cost function is not quadratic in the parameters. For parameter estimation of dynamical systems, this measure is intrinsically different from the measure used in the SPS method. Analysing the behaviour of this performance measure is an intriguing prospect. The membership test in this case would require solution of the estimation problems corresponding to the perturbed datasets, which is computationally expensive. Also analysing the structure of the corresponding confidence regions becomes much more difficult.

The assumed properties of the noise pretty much determine the range of possible perturbations (for the two presented cases there are no other options than the ones used). As a direct consequence the only really tunable point of the framework is the performance measure. 

Finally, maybe the most important takeaway message of the presented analysis is that thorough analysis should be carried out regarding the structure of the confidence regions corresponding to the chosen performance measure and the selected orderings that are accepted. If confidence regions are visualized, guaranties are required that the entire region is shown, otherwise the visualized subset might not have the required exact confidence level.

\begin{ack}                               
This work has been supported partly by the MTA-BME Control Engineering Research Group of the HAS at the Budapest University of Technology and Economics. Support was also provided by the Fund for Scientific Research (FWO-Vlaanderen), in part by the Flemish Government (Methusalem) and in part by the Belgian Government through the Interuniversity Poles of Attraction (IUAP VI/4) Program.
\end{ack}

\bibliographystyle{plain}        
\bibliography{kolumban}           

\begin{thebibliography}{10}

\bibitem{Aldous1985}
David~J. Aldous.
\newblock {\em Exchangeability and related topics}.
\newblock Lecture Notes in Mathematics. Springer Berlin Heidelberg, 1985.

\bibitem{Campi2012Plenary}
M.~C. Campi, B.~Cs. Cs\'aji, S.~Garatti, and E.~Weyer.
\newblock Certified system identification: Towards distribution-free results.
\newblock In {\em Proceedings of the 16th IFAC Symposium on System
  Identification (SYSID 2012)}, pages 245--255, 2012.

\bibitem{Campi2005NonAssID}
M.C. Campi and E.~Weyer.
\newblock Guaranteed non-asymptotic confidence regions in system
  identification.
\newblock {\em Automatica}, 41(10):1751 -- 1764, 2005.

\bibitem{Campi2010NonAssLinearTF}
M.C. Campi and E.~Weyer.
\newblock Non-asymptotic confidence sets for the parameters of linear transfer
  functions.
\newblock {\em Automatic Control, IEEE Transactions on}, 55(12):2708--2720,
  2010.

\bibitem{Csaji2012Sysid}
B.Cs. Cs\'aji, M.C. Campi, and E.~Weyer.
\newblock Non-asymptotic confidence regions for the least-squares estimate.
\newblock In {\em Proceedings of the 16th IFAC Symposium on System
  Identification (SYSID 2012)}, pages 227--232, 2012.

\bibitem{Csaji2012CDC}
B.Cs. Cs\'aji, M.C. Campi, and E.~Weyer.
\newblock Sign-perturbed sums (sps): A method for constructing exact
  finite-sample confidence regions for general linear systems.
\newblock In {\em Decision and Control (CDC), 2012 IEEE 51st Annual Conference
  on}, pages 7321--7326, 2012.

\bibitem{Dalai2007LSCRNonLin}
Marco Dalai, Erik Weyer, and Marco~C. Campi.
\newblock Parameter identification for nonlinear systems: Guaranteed confidence
  regions through lscr.
\newblock {\em Automatica}, 43(8):1418 -- 1425, 2007.

\bibitem{Garatti2004}
S.~Garatti, M.C. Campi, and S.~Bittanti.
\newblock Assessing the quality of identified models through the asymptotic
  theory - when is the result reliable?
\newblock {\em Automatica}, 40(8):1319 -- 1332, 2004.

\bibitem{Kieffer2014507}
Michel Kieffer and Eric Walter.
\newblock Guaranteed characterization of exact non-asymptotic confidence
  regions as defined by \{LSCR\} and \{SPS\}.
\newblock {\em Automatica}, 50(2):507 -- 512, 2014.

\bibitem{ljung1999SysId}
Lennart Ljung, editor.
\newblock {\em System Identification (2Nd Ed.): Theory for the User}.
\newblock Prentice Hall PTR, Upper Saddle River, NJ, USA, 1999.

\end{thebibliography}




\end{document}